\documentclass[journal]{IEEEtran}
\usepackage{graphicx}
\usepackage{amssymb}
\usepackage{multirow}
\usepackage{indentfirst}
\usepackage{inputenc}
\usepackage{array}
\usepackage{multicol}
\usepackage[switch,mathlines]{lineno}
\usepackage{soul}
\usepackage{graphicx}
\usepackage{cite}
\usepackage[cmex10]{amsmath}
\usepackage{algorithmic}
\usepackage{array}
\usepackage{CJK}
\usepackage[cmex10]{amsmath}
\usepackage{algorithmic}
\usepackage{array}
\usepackage{fixltx2e}
\usepackage{bm}
\usepackage{amssymb}
\usepackage{color}
\usepackage{nomencl}
\usepackage{setspace}
\usepackage[ruled, linesnumbered]{algorithm2e}
\ifCLASSOPTIONcompsoc
 \usepackage[caption=false,font=normalsize,labelfont=sf,textfont=sf]{subfig}
\else
 \usepackage[caption=false,font=footnotesize]{subfig}
\fi
\usepackage{url}
\usepackage{makecell}
\usepackage{colortbl}
\usepackage{color}
\usepackage[table,xcdraw]{xcolor}
\def\BibTeX{{\rm B\kern-.05em{\sc i\kern-.025em b}\kern-.08em
    T\kern-.1667em\lower.7ex\hbox{E}\kern-.125emX}}
\usepackage{balance}
\usepackage{booktabs, tabularx}
\usepackage[referable]{threeparttablex}
\usepackage{stfloats}
\usepackage{multirow}
\usepackage{float}
\usepackage{hyperref}
\hypersetup{
    colorlinks=True,
    pdfborder={0 0 0},
    linkcolor=blue,    
    citecolor=blue,    
    filecolor=magenta,  
    menucolor=red,      
    urlcolor=cyan,      
}
\begin{document}

\title{Comparative Study of Data-driven Area Inertia Estimation Approaches on WECC Power Systems}

\author{{Bendong Tan,  Jiangkai Peng, Ningchao Gao, Junbo~Zhao, Jin Tan}
  \thanks{This work was authored in part by the National Renewable Energy Laboratory, operated by Alliance for Sustainable Energy, LLC, for the U.S. Department of Energy (DOE) under Contract No. DE-AC36-08GO28308. This material is based upon work supported by the U.S. Department of Energy's Office of Energy Efficiency and Renewable Energy (EERE) under the Solar Energy Technologies Office Award Number 37772. B. Tan and J. Zhao are with University of Connecticut, Storrs, CT 06269. This work was based on Bendong Tan's summer internship at NREL. J. Peng, N. Gao and J. Tan are with National Renewable Energy Laboratory, Golden, CO 80401. (e-mail:  jin.tan@nrel.gov)}}

\markboth{IEEE Power Energy Society General Meeting, 2024}%
{Shell \MakeLowercase{\textit{et al.}}: Bare Demo of IEEEtran.cls for Journals}

\maketitle
\begin{abstract}
  With the increasing integration of inverter-based resources into the power grid, there has been a notable reduction in system inertia, potentially compromising frequency stability. To assess the suitability of existing area inertia estimation techniques for real-world power systems, this paper presents a rigorous comparative analysis of system identification, measurement reconstruction, and electromechanical oscillation-based area inertia estimation methodologies, specifically applied to the large-scale and multi-area WECC 240-bus power system. Comprehensive results show that the system identification-based approach exhibits superior robustness and accuracy relative to its counterparts.

\end{abstract}

\vspace{-0.4cm}
\section{Introduction}
The role of inertia in ensuring the frequency stability of power systems is pivotal \cite{Kundur_1994}. Nonetheless, the substantial integration of inverter-based resources (IBRs), such as wind energy and solar, has led to a marked decline in system inertia.  Consequently, obtaining an accurate understanding of inertia distribution becomes imperative for the secure operation of power systems \cite{Tan_2022}.

Extensive research has been dedicated to inertia estimation at the system level, focusing on quantifying the cumulative inertia within an entire power system. This approach conceptualizes all generators as a singular equivalent entity, the dynamics of which are articulated through the aggregated swing equation, framed under the center of inertia (COI) coordinate system \cite{Kundur_1994}. Regarding the system-level inertia estimation, both the total disturbed active power and the COI frequency are requisite \cite{Wall_2018}. System inertia can be estimated via the dynamics of the aggregated swing equation or the definition of inertia. The COI frequency can be depicted as the weighted average frequency derived from phasor measurement units (PMUs) positioned at various locations \cite{Schiffer_2019,Wang_2019, Phurailatpam_2019}. On the other hand, the total disturbed power can be obtained via recorded large disturbance \cite{Wang_2019} or the total power variation at each generator \cite{Makolo_2021}. 

However, acquiring the total disturbed power poses challenges, primarily attributed to the complexities in disturbance power quantification and insufficient PMU placements. Furthermore, in the context of systems with a significant presence of IBRs, inertia might be unevenly distributed throughout the system \cite{Zeng_2020}. This introduces spatial variances, nuances that system-level inertia estimation fails to capture. While the overall system inertia might seem satisfactory, specific regions may experience an inertia deficit, potentially giving rise to stability concerns.

Hence, if system operators are inclined towards understanding inertia distribution, area inertia estimation emerges as a more favorable approach. Analogous to system-level inertia estimation, area inertia determination relies on the total disturbed power within the area and the corresponding area COI frequency. However, the cumulative disturbed power within a specific area can be ascertained by aggregating the changes in tie-line power \cite{Li_2018}. Approaches for area inertia estimation can be categorized into three distinct categories: (1) System identification-based methods; (2) Measurement reconstruction-based methods; (3) Electromechanical oscillation-based methods. Within the realm of system identification-based methodologies, both \cite{Tuttelberg_2018} and \cite{Zeng_2020} employ a reduced-order model, i.e., the transfer function, to emulate the dynamic input-output relationship between the total disturbed area power and the area COI frequency. This results in a straightforward computation of area inertia, anchored on the inertia's fundamental definition. Transitioning to the measurement reconstruction-based techniques, \cite{Phurailatpam_2020} deploys polynomial expansion to reconstruct both power deviation and frequency. Concurrently, \cite{Yang_2021} harnesses dynamic mode decomposition (DMD), yielding an analytical framework for deducing area inertia. Delving into the electromechanical oscillation-based strategies, both \cite{Cai_2019} and \cite{Wang_2022} establish the relationship between the electromechanical oscillation mode and area inertia, facilitating the quantification of the latter.

The previously discussed inertia estimation methodologies, tested on benchmark systems with coherent generator dynamics, have yet to be comprehensively compared in the same benchmark system or real-world systems under practical testing scenarios where areas are not strictly defined by generator coherency. The goal of this paper is to understand the pros and cons of each method, i.e, on the WECC 240-bus system, a reduced model of the US western grid. Here, frequency stability is maintained through dispatch in balanced areas, which may not always correspond to generator coherency \cite{Yuan_2020, WECC_2020}.
\vspace{-0.4cm}
\section{Problem Formulation}
\begin{figure}[!t]
\centering
\includegraphics[width=1.6in]{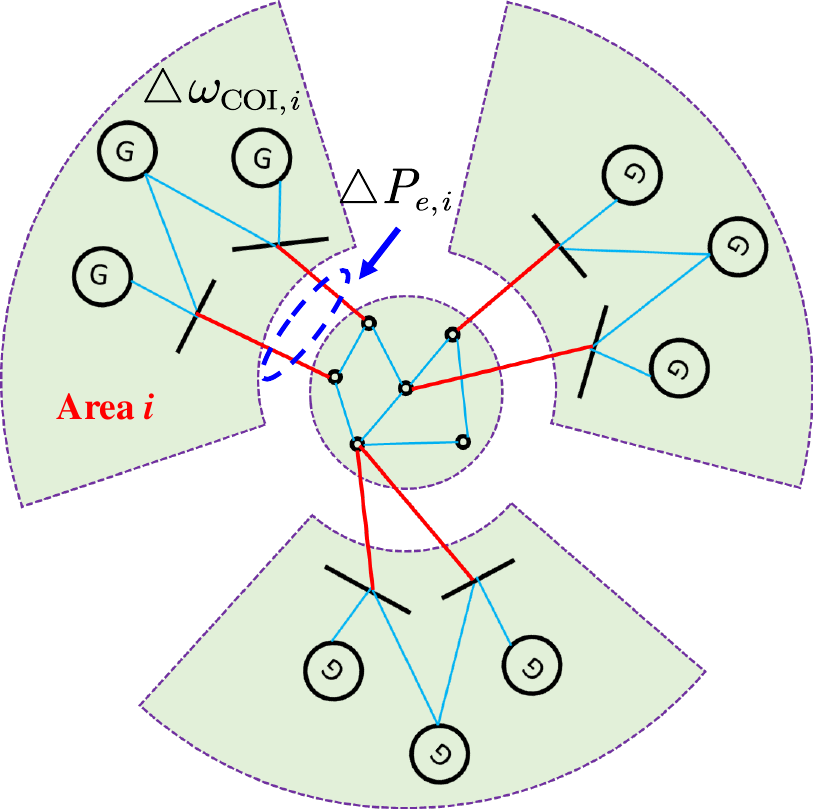}
\caption{Diagram of the multi-area power system.}
\label{fig_regional_power_system}
\end{figure}

In a specific regional power system, as shown in Fig. \ref{fig_regional_power_system}, the dynamics between the area's tie-line power and its center of inertia (COI) frequency can be effectively characterized by representing the system with an equivalent synchronous generator. This representation yields an aggregated swing equation given by:
\begin{equation}
\dot{\omega}_{\text{COI},i} = \frac{1}{2 H_{i}}\left(P_{m, i}-P_{e, i}-D_{i} \Delta \omega_{\text{COI},i}\right)
\label{eq_equivalent_swing_equation}
\end{equation}
where $\omega_{\text{COI},i}$ denotes the COI frequency for the $i$-th area: 
\begin{equation}
\omega_{\text{COI},i}={{\underset{m \in \text{Area} \: i}{\sum} \omega_{m} \text{S}_{m}}\Bigg/{\underset{m \in \text{Area} \: i}{\sum} \text{S}_{m}}}
\end{equation}
where $\omega_{m}$ and $\text{S}_{m}$ are the generator m's rotor speed and rated power respectively.  $P_{m, i}$ and $P_{e, i}$ are representative of the tie-line total mechanical and electrical power for the $i$-th area, respectively. Additionally, $H_{i}$ signifies the equivalent inertia and $D_{i}$ the damping factor pertinent to the $i$-th area. Since primary frequency control can be formulated as  $K_{i} \Delta \omega_{\text{COI},i}$, where $K_{i}$ is the primary frequency control coefficient, the perturbations effects associated with $P_{m, i}$ can be included into term $D_{i} \Delta \omega_{\text{COI},i}$ in (\ref{eq_equivalent_swing_equation}). Therefore, (\ref{eq_equivalent_swing_equation}) can be further reformulated in an incremental form:
\begin{equation}
\Delta \dot{\omega}_{\text{COI},i} = \frac{1}{2 H_{i}}\left(-\Delta P_{e, i}-D_{i} \Delta \omega_{\text{COI},i}\right)
\label{eq_incremental_swing_equation}
\end{equation}
where $\Delta P_{e, i}$ signifies the deviation in electrical power from its steady-state value.  
When $P_{m, i}$ undergoes dynamic changes, as observed during the initiation of primary frequency control, (\ref{eq_incremental_swing_equation}) yields a higher estimate for the equivalent damping factor.

To estimate \( H_{i} \), the primary challenge lies in accurately determining \( \Delta \dot{\omega}_{\text{COI},i} \). Relying solely on the finite difference method for frequency measurements can introduce numerical errors. Therefore, existing state-of-art methods try to estimate inertia without directly calculating \( \Delta \dot{\omega}_{\text{COI},i} \).
\vspace{-0.4cm}
\section{Area Inertia Estimation Approaches}
In the following section, we benchmark cutting-edge methodologies from each category of area inertia estimation on a large-scale practical power system. Specifically, we adopt the techniques proposed in \cite{Zeng_2020}, \cite{Yang_2021}, and \cite{Wang_2022} which represent the system identification-based, measurement reconstruction-based, and electromechanical oscillation-based methods, respectively. It is worth mentioning that enhancements have been incorporated into the techniques outlined in \cite{Zeng_2020} and \cite{Yang_2021} to address and mitigate numerical challenges.
\vspace{-0.3cm}
\subsection{System Identification-based Method}
The Laplace transformation applied to both sides of (\ref{eq_incremental_swing_equation}) results in the relation $\frac{\Delta \omega_{\text{COI}, i}(s)}{\Delta P_{e, i}(s)} \approx-\frac{1}{2 H_{i} s+D_{i}}$. Acknowledging the influences of additional control components linked with the generator, the central premise of \cite{Zeng_2020} is to employ a higher-order transfer function, denoted as $G_{i}(s)$, to act as a surrogate model elucidating the relationship between $\omega_{\text{COI}, i}$ and $\Delta P_{e, i}$. The form of $G_{i}(s)$ is given by:
\begin{equation}
G_{i}(s) = \frac{1}{a_{n} s^{n}+\cdots +a_{1} s+a_{0}},
\label{eq_transfer_function}
\end{equation}
where $a_{n}$ signifies the coefficient of the $n$th-order term. Given the presence of other control modules linked to generators, \( n = 2 \) is deemed sufficient to approximate the relationship between \( \omega_{\text{COI}, i} \) and \( \Delta P_{e, i} \). The parameters intrinsic to $G_{i}(s)$ can be discerned through the N4SID (Numerical Algorithm for Subspace State Space System Identification) method, leveraging the input $\Delta P_{e, i}$ and output $\Delta \omega_{\text{COI}, i}$ \cite{Zeng_2020}.

Subjecting $G_{i}(s)$ to a step function (emulating a scenario where $\Delta P_{e, i}=1$ pu) yields the step response $g_{\text{step}}(t)$. This step response can be approximated using an $N_{p}$-order polynomial:
\begin{equation}
g_{\text{step}}(t) = c_{N_{p}}t^{N_{p}}+\cdots +c_{1} t+c_{0}
\end{equation}
where $t$ is the temporal variable and $c_{N_{p}}$ is the coefficient of the $N_{p}$-order term of $g_{\text{step}}(t)$. Given the small magnitude of $D_{i}$, which can be approximated as negligible, we propose to estimate $H_{i}$ with (\ref{eq_inertia_est}):
\begin{equation}
\begin{split}
\widetilde{H}_{i} &=-\left. \frac{\Delta P_{e, i}(t)}{2 \dot{\omega}_{\text{COI}, i}(t)} \right|_{t=0} =-\left. \frac{1}{2\dot{g}_{\text{step}}(t)} \right|_{t=0} = - \frac{1}{2c_{1}}\\
\end{split}
\label{eq_inertia_est}
\end{equation}
where $\widetilde{H}_{i}$ is estimated equivalent inertia for the $i$-th area. In
the swing equation, inertia values of synchronous generator
range from 1.75 s to approximately 10 s, and the damping
factor varies from the magnitude of $10^{-2}$ to $10^{-1}$ \cite{Tan_2022, Panda_2020}.
Hence, the damping effect is small enough to be neglected.
The exclusion of \( D_{i} \) in expression (6) is expected to result in a slightly biased estimate of inertia. Generators with similar rotor angle response following a disturbance are called
coherent generators. The coherency between generators, particularly over large electrical distances, may not be significant, potentially leading to substantial errors in \( \omega_{\text{COI}} \) as an aggregate frequency representation. However, since the system identification-based method solely utilizes dynamics at the initial stage ($t=0$), the coherency of area generators exerts minimal influence on its performance.
\vspace{-0.4cm}
\subsection{Measurement Reconstruction-based Method}
A typical approach for measurement reconstruction-based method is using DMD to reconstruct the measurements of $\boldsymbol{X} \in \mathbb{R}^{2N_{a} \times M} $, where $\boldsymbol{X} = [\Delta 
 \boldsymbol{\omega}_{\text{COI}, 1},\; \Delta 
  \boldsymbol{\omega}_{\text{COI}, 2} \; ,\cdots, \Delta 
  \boldsymbol{\omega}_{\text{COI}, N_{a}}\;,\Delta   \boldsymbol{P}_{e, 1}\;\Delta \boldsymbol{P}_{e, 2}\;\cdots,\Delta \boldsymbol{P}_{e, N_{a}}]^{\mathsf{T}}$  \cite{Yang_2021}. $N_{a}$ is the number of areas in the power system, while $M$ is the number of measurements.
\begin{equation}
      A_{\boldsymbol{c}}\left[\begin{array}{c}
\boldsymbol{\varphi}_{k}^{\Delta \boldsymbol{\omega}_{\text{COI}}} \\
-- \\
\boldsymbol{\varphi}_{k}^{\Delta \boldsymbol{P}_{e}}
\end{array}\right]=\lambda_{k}\left[\begin{array}{c}
\boldsymbol{\varphi}_{k}^{\Delta \boldsymbol{\omega}_{\text{COI}}} \\
-- \\
\boldsymbol{\varphi}_{k}^{\Delta \boldsymbol{P}_{e}}
\end{array}\right]
\label{eq_eigen_decomposition}
  \end{equation}
where $A_{\boldsymbol{c}} = \text{ln}(A_{\boldsymbol{d}})/ \Delta t$;  
$\Delta t$ is the sampling time interval; $A_{\boldsymbol{d}}$ is the state transition matrix that leads to $\boldsymbol{X}_{2:M} = A_{\boldsymbol{d}}\boldsymbol{X}_{1:M-1}$; $\boldsymbol{X}_{2:M}$ denotes the measurements from time instants 1 to $M-1$; $\boldsymbol{\varphi}_{k}^{\Delta \boldsymbol{P}_{e}}$ and $\boldsymbol{\varphi}_{k}^{\Delta \boldsymbol{P}_{e}}$ denote the the $k$-th DMD modes with respect to
$\Delta \boldsymbol{\omega}_{\text{COI}}$ and $\Delta \boldsymbol{P}_{e}$; $\lambda_{k}$ is the $k$-th eigenvalue. As a result, measurements can be reconstructed as:
\begin{equation}
    \left[\begin{array}{c}
\Delta \boldsymbol{\omega}_{\text{COI}} \\
\Delta \boldsymbol{P}_{e}
\end{array}\right]=\sum_{k=1}^{2N_{a}} e^{\lambda_{k} t} b_{k}\left[\begin{array}{c}
\boldsymbol{\varphi}_{k}^{\Delta \boldsymbol{\omega}_{\text{COI}}} \\
\boldsymbol{\varphi}_{k}^{\Delta \boldsymbol{P}_{e}}
\end{array}\right]
\label{eq_reconstruction}
\end{equation}
where $b_{k}$ is the initial value coefficient corresponding to
$k$-th eigenvalues. By applying (\ref{eq_reconstruction}) to (\ref{eq_incremental_swing_equation}), (\ref{eq_inertia_est_dmd}) can be derived:
\begin{equation}
\left[ \begin{matrix}
	\mathrm{Re}\left( \varphi ^{\Delta \omega_{\text{COI}}}\lambda \boldsymbol{b} \right)&		\mathrm{Re}\left( \varphi ^{\Delta \boldsymbol{\omega}_{\text{COI}}}\boldsymbol{b} \right)\\
	\mathrm{Im}\left( \varphi ^{\Delta \omega_{\text{COI}}}\lambda \boldsymbol{b} \right)&		\mathrm{Im}\left( \varphi ^{\Delta \boldsymbol{\omega}_{\text{COI}}}\boldsymbol{b} \right)\\
\end{matrix} \right] \left[ \begin{array}{c}
	2\boldsymbol{\widetilde{H}}\\
	\boldsymbol{D}\\
\end{array} \right] =-\left[ \begin{array}{l}
	\mathrm{Re}\left( \varphi ^{\Delta \boldsymbol{P}_{e}}\boldsymbol{b} \right)\\
	\mathrm{Im}\left( \varphi ^{\Delta \boldsymbol{P}_{e}}\boldsymbol{b} \right)\\
\end{array} \right]
\label{eq_inertia_est_dmd}
\end{equation}

In the work of \cite{Yang_2021}, inertia is inferred through the resolution of (\ref{eq_inertia_est_dmd}). Nonetheless, our observations suggest potential inaccuracies in scenarios where the aforementioned equation admits multiple solutions. In an attempt to constrain and refine the solution space, we introduce additional constraints:
\begin{equation}
   0 \le \boldsymbol{D} \le 1
   \label{eq_dmd_constrains}
\end{equation}
By jointly solving (\ref{eq_inertia_est_dmd}) and (\ref{eq_dmd_constrains}), we posit that the inertia estimation converges to a more pragmatic and precise value.

\vspace{-0.4cm}
\subsection{Electromechanical Oscillation-based Method}
The study by \cite{Wang_2022} establishes a connection between area inertia and electromechanical oscillations, facilitating the inference of area inertia directly from these oscillations. Taking into account the frequency bandwidth \(B\) of the electromechanical oscillation, the Fourier transform of (\ref{eq_incremental_swing_equation}) can be derived as:
\begin{equation}
    2H_{i} \sum_{\gamma=0}^{\gamma=B} \Delta \dot{\omega}_{\text{COI},i}(\gamma)+D_{i} \sum_{\gamma=0}^{\gamma=B} \Delta \omega_{\text{COI},i}(\gamma)+\sum_{\gamma=0}^{\gamma=B} \Delta P_{e, i}(\gamma)=0
    \label{eq_fft}
\end{equation}
where $\gamma$ is the frequency in the oscillation band. According to (\ref{eq_fft}), $H_{i}$ can be estimated via (\ref{eq_fft_inertia_est}).

\begin{figure*}[b]
\begin{equation}
\widetilde{H}_{i}=\frac{1}{2}\frac{\operatorname{Re}\left[\sum_{\gamma=0}^{\gamma=B} \Delta \omega_{\text{COI},i}(\gamma)\right] \operatorname{Im}\left[\sum_{\gamma=0}^{\gamma=B} \Delta P_{e, i}(\gamma)\right]-\operatorname{Re}\left[\sum_{\gamma=0}^{\gamma=B} \Delta P_{e, i}(\gamma)\right] \operatorname{Im}\left[\sum_{\gamma=0}^{\gamma=B} \Delta \omega_{\text{COI},i}(\gamma)\right]}{\operatorname{Re}\left[\sum_{\gamma=0}^{\gamma=B} j \gamma \Delta \omega_{\text{COI},i}(\gamma)\right] \operatorname{Im}\left[\sum_{\gamma=0}^{\gamma=B} \Delta \omega_{\text{COI},i}(\gamma)\right]-\operatorname{Re}\left[\sum_{\gamma=0}^{\gamma=B} \Delta \omega_{\text{COI},i}(\gamma)\right] \operatorname{Im}\left[\sum_{\gamma=0}^{\gamma=B} j \gamma \Delta \omega_{\text{COI},i}\right]}
\label{eq_fft_inertia_est}
\end{equation}
\end{figure*}
\vspace{-0.4cm}
\subsection{Comparisons among Various Methods}
From the comprehensive analysis of the methodologies presented, it is evident that none necessitate the computation of COI frequency derivatives, a process known to encounter numerical challenges, particularly in the presence of noise or outliers.

(1) The pros of each method are listed as follows:
\begin{itemize}
    \item \textbf{System Identification:} This method offers ease of use due to its limited hyperparameters, which minimally impact its performance.

    \item \textbf{Measurement Reconstruction:} Beyond inertia estimation, this method also furnishes damping estimation, thereby yielding additional dynamic information.
    \item \textbf{Electromechanical Oscillation:} Similar to measurement reconstruction-based method, this method can also provide damping estimation.

\end{itemize}

(2) The cons of each method and corresponding measures are summarized as follows:
\begin{itemize}
    \item \textbf{System Identification:} Despite yielding a biased estimation due to the omission of the damping factor, the resultant error is deemed acceptable, considering the relatively minor magnitude of the damping factor.
    \item \textbf{Measurement Reconstruction:} Given that \( b \), the initial value coefficient associated with eigenvalues in measurement reconstruction, is determined by the first data point in response measurements, noise considerably influences the final inertia estimation. Consequently, noise processing and careful choice of the start point of the response measurements is essential.
    \item \textbf{Electromechanical Oscillation:} As delineated in (\ref{eq_fft_inertia_est}), inertia estimation accuracy is contingent upon the bandwidth \( B \), underscoring its sensitivity to \( B \). Thus, precise estimation of \( B \) is crucial.

\end{itemize}
\begin{figure}[htb]
\centering
\includegraphics[width=3.6in]{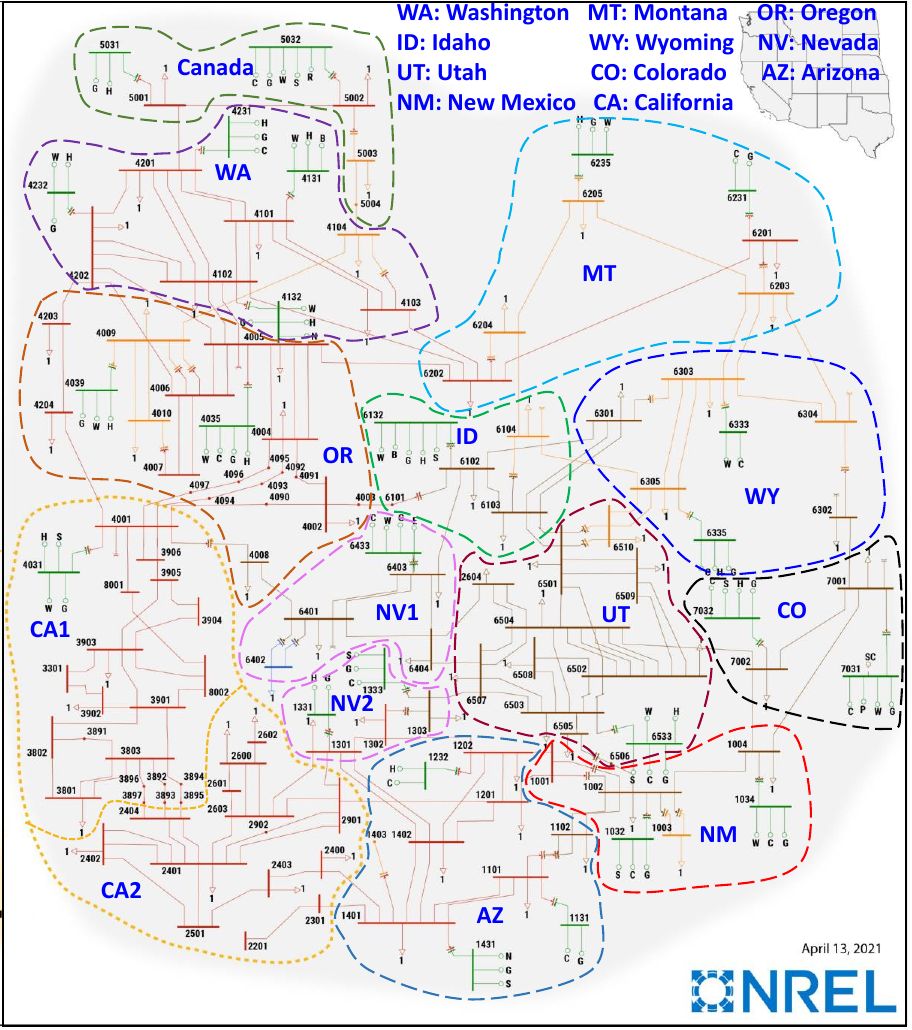}
\caption{One-line diagram of WECC 240-bus power system.}
\label{fig_wecc}
\end{figure}
\vspace{-0.4cm}
\section{Numerical Results}
To comprehensively compare the pros and cons of each method, extensive validations have been conducted on the WECC 240-bus power system, where a total of 13 regional power grids have been delineated according to balanced areas, as shown in Fig. \ref{fig_wecc}. Every scenario was examined in the presence of a disturbance, characterized by a three-phase short-circuit fault at Bus 1002, initiated at \( t = 1 \)s and cleared by \( t = 1.1 \)s. The entirety of the simulation spanned 10 seconds.

\vspace{-0.4cm}
\subsection{Parameter Sensitivity Analysis for Different Methods}
\begin{figure}[h]
\centering
\subfloat[]{\includegraphics[width=3in]{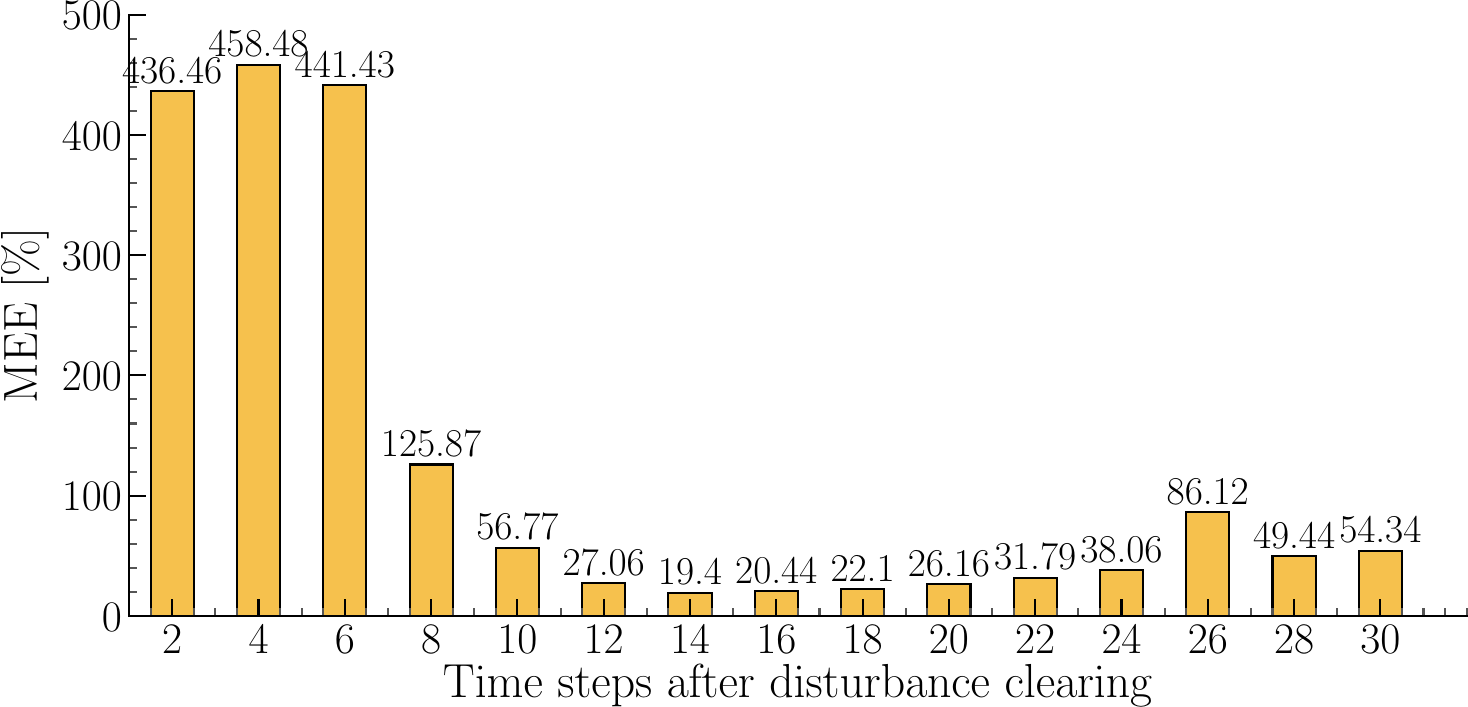}
}
\vfil
\subfloat[]{\includegraphics[width=1.74in]{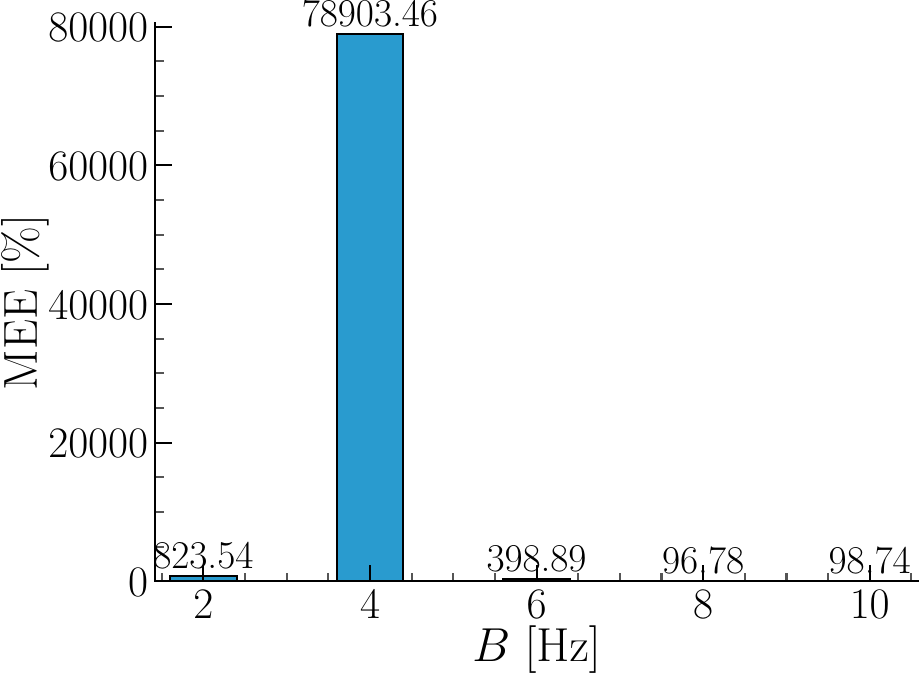}%
}
\hfil
\subfloat[]{\includegraphics[width=1.74in]{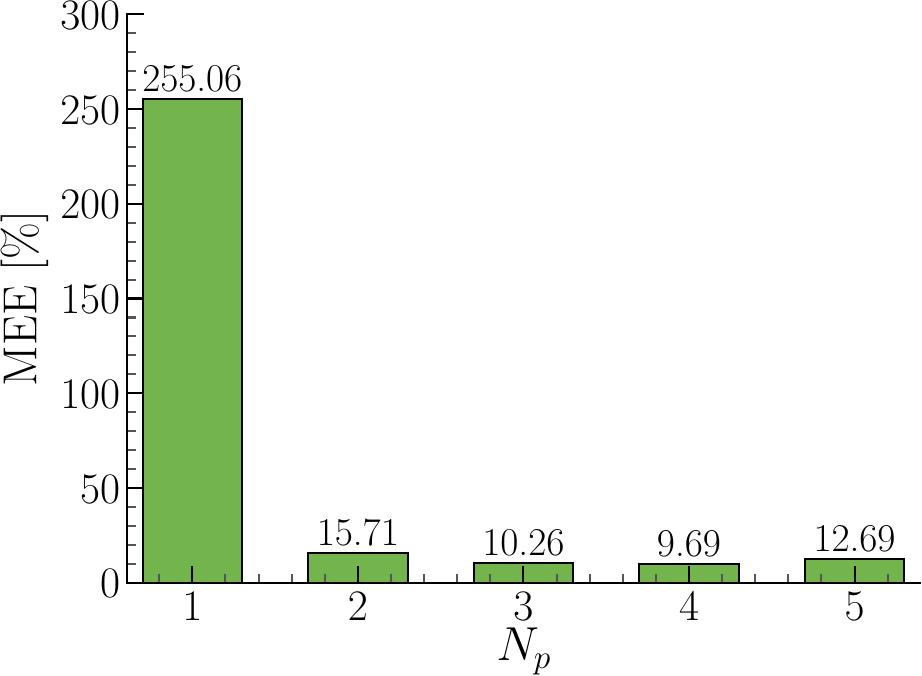}%
}
\caption{Maximum area inertia estimation error under different hyperparameters for various methods. (a) Measurement reconstruction-based method; (b) System identification-based method; (c) Electromechanical oscillation-based method.}
\label{fig_parameter_sensitivity}
\end{figure}
Upon reviewing various methodologies, it becomes evident that the polynomial order \(N_{p}\), the initial value \(\boldsymbol{b}\), and the oscillation bandwidth \(B\) serve as hyperparameters for system identification, measurement reconstruction, and the electromechanical oscillation-based method, respectively. For \(\boldsymbol{b}\), it is imperative to designate the specific instant at which at which the starting point of the measurement response is selected. To evaluate the efficacy of each method, we introduce two indices: the maximum estimation error (MEE) and the estimation error (EE):
\begin{equation}
    \text{MEE} = \underset{1\le i \le N_{a}}{\text{max}} \left(\left | \frac{\widetilde{H}_{i}-H_{i}}{H_{i}}  \right |   \times 100\%\right) \: , \: \text{EE} = \left | \frac{\widetilde{H}_{i}-H_{i}}{H_{i}}  \right |  \times 100\%
\end{equation}

From Fig. \ref{fig_parameter_sensitivity}, it is evident that an increased \(N_{p}\) enhances the fitting quality for \(g_{\text{step}(t)}\), thereby improving the accuracy of area inertia estimation. Specifically, \(N_{p} = 4\) emerges as an optimal choice, yielding the minimal MEE for the system identification-based method. For the measurement reconstruction-based method, the initial values \(\boldsymbol{b}\) are chosen at several time steps post-disturbance clearing. As depicted in Fig. \ref{fig_parameter_sensitivity}(a), selecting \(\boldsymbol{b}\) shortly after disturbance clearing results in a MEE exceeding 400\%. This can be attributed to the fact that while DMD offers a linear approximation of the dynamic system, measurements taken shortly post-disturbance exhibit pronounced nonlinear characteristics. Consequently, \(\boldsymbol{b}\) is determined using the measurement from the 14-th time step post-disturbance to minimize the estimation error. Regarding the electromechanical oscillation-based method, its elevated MEE, reaching up to 100\%, stems from the formulation of (\ref{eq_fft_inertia_est}) through the continuous domain's Fourier transform. This necessitates approximation via the discrete Fourier transform, given that measurements are discretely acquired \cite{Wang_2022}. Owing to the substantial estimation error associated with the electromechanical oscillation-based method, it is excluded from subsequent comparative analyses.
\vspace{-0.4cm}
\subsection{Area Inertia Estimation Results Under Noise}

\begin{figure}[!t]
\centering
\subfloat[]{\includegraphics[width=3in]{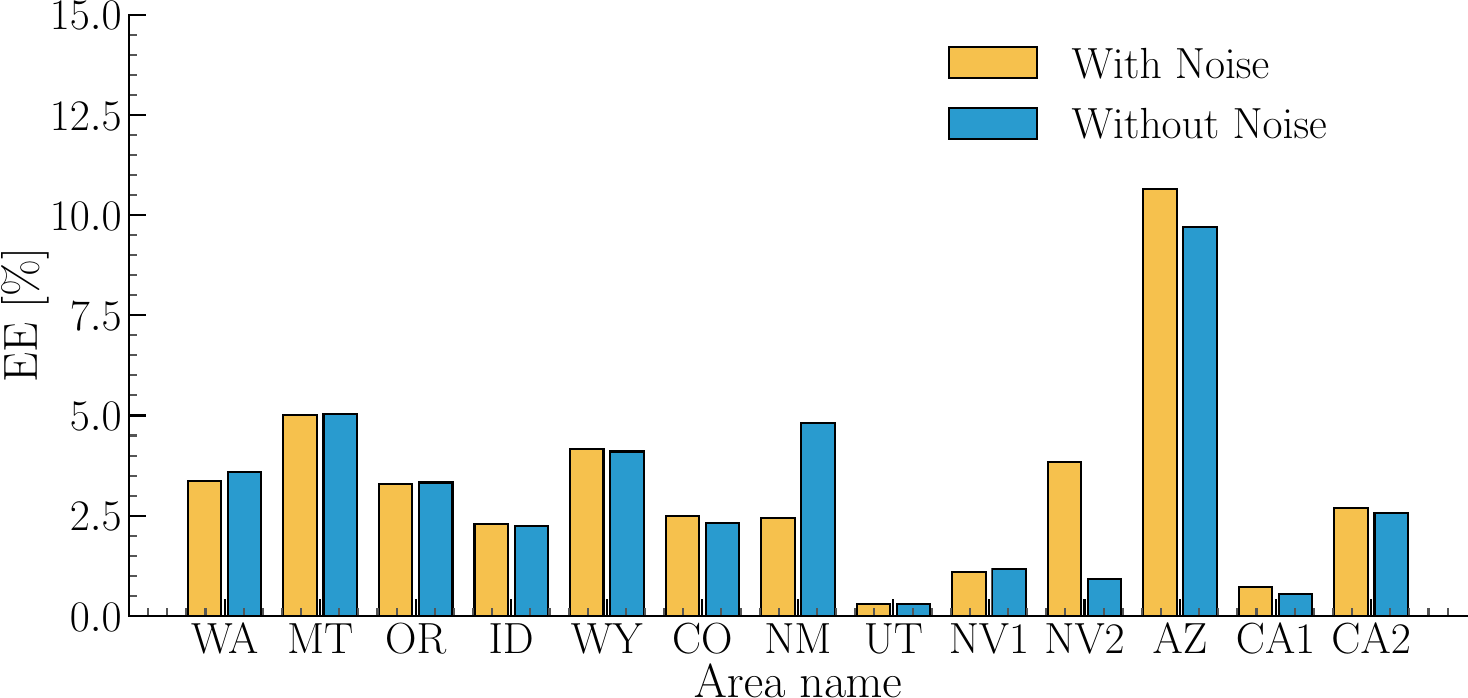}
}
\vfil
\subfloat[]{\includegraphics[width=3in]{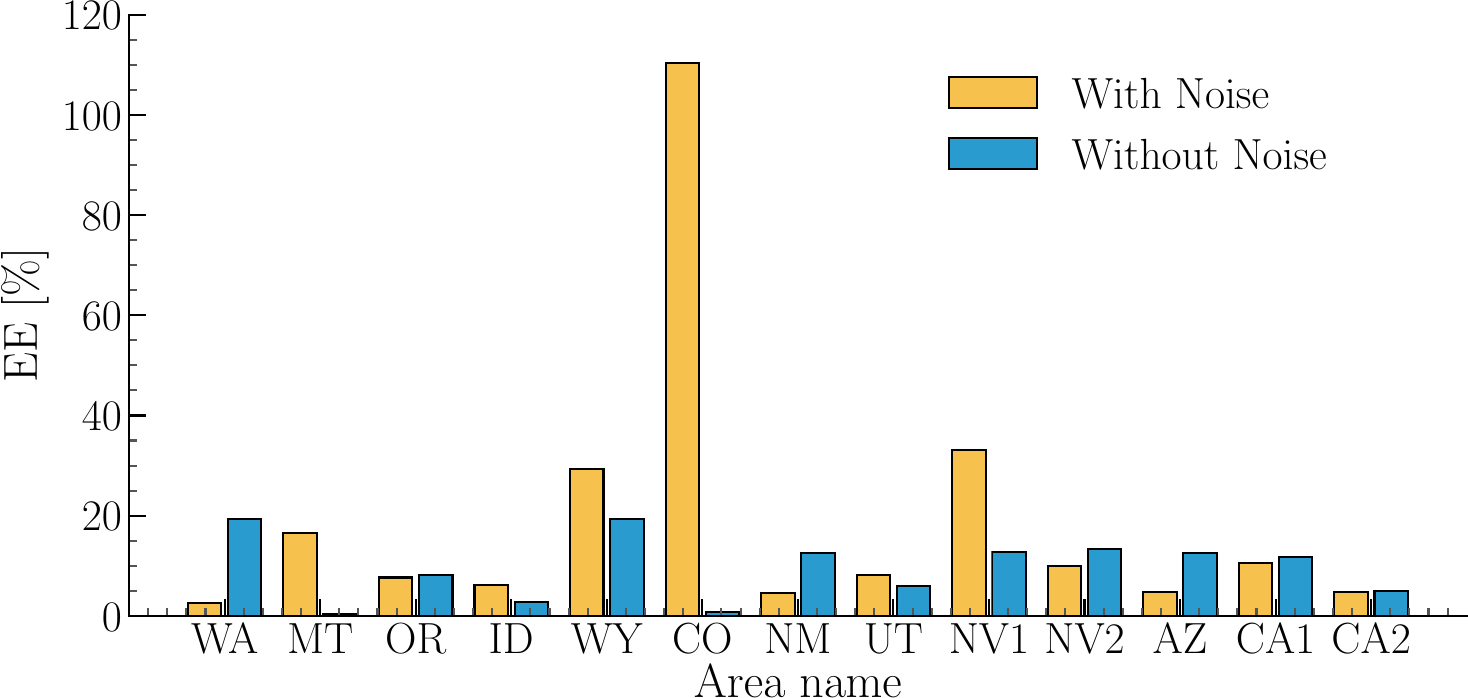}
}
\caption{Area inertia estimation error for various methods. (a) System identification-based method; (b) Measurement reconstruction-based method.}
\label{fig_noise_performance}
\end{figure}

In this section, the performance of system identification and measurement reconstruction-based methods are validated under measurement noise. The measurement noise is postulated to follow a zero-mean Gaussian distribution with a standard deviation of \(\sigma = 4 \times 10^{-3}\) \cite{Singh_2009}. Notably, this is greater than the realistic noise encountered in measurement data, which has an upper limit of 1\% total vector error (TVE). To mimic the data preprocessing typically encountered in practical operations, noisy measurements undergo smoothing using a low-pass filter with a cut-off frequency of 5Hz \cite{Zeng_2020}. Fig. \ref{fig_noise_performance} illustrates the inertia estimation errors for each area as determined by both the system identification and measurement reconstruction-based methods. In both noise-free and noisy conditions, the system identification-based method consistently yields estimation errors below 5\% for all areas, with the exception of AZ. This underscores the method's robustness. In contrast, the measurement reconstruction-based method exhibits area inertia estimation errors exceeding 100\%. Such discrepancies might arise from noise-induced alterations in the initial values \(\boldsymbol{b}\), which in turn precipitate significant estimation errors.
\vspace{-0.4cm}
\subsection{Estimation Results under Different Operation Conditions}
\begin{figure}[!t]
\centering
\includegraphics[width=3.2in]{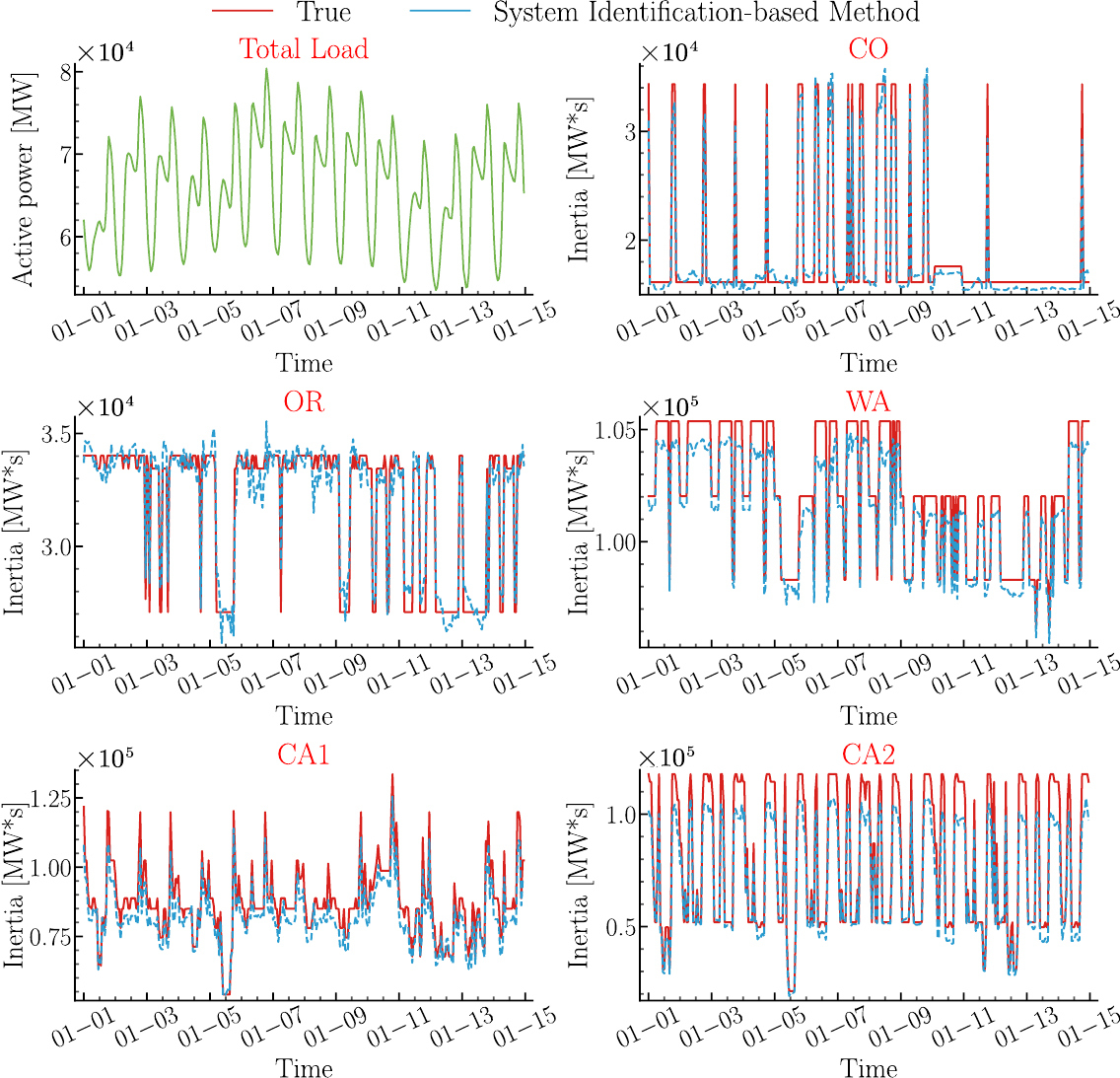}
\caption{Inertia estimation results of system identification-based method for each area from January 1st to January 14th, 2019.}
\label{fig_transfer_inertis_est}
\end{figure}
\begin{figure}[!t]
\centering
\includegraphics[width=3.2in]{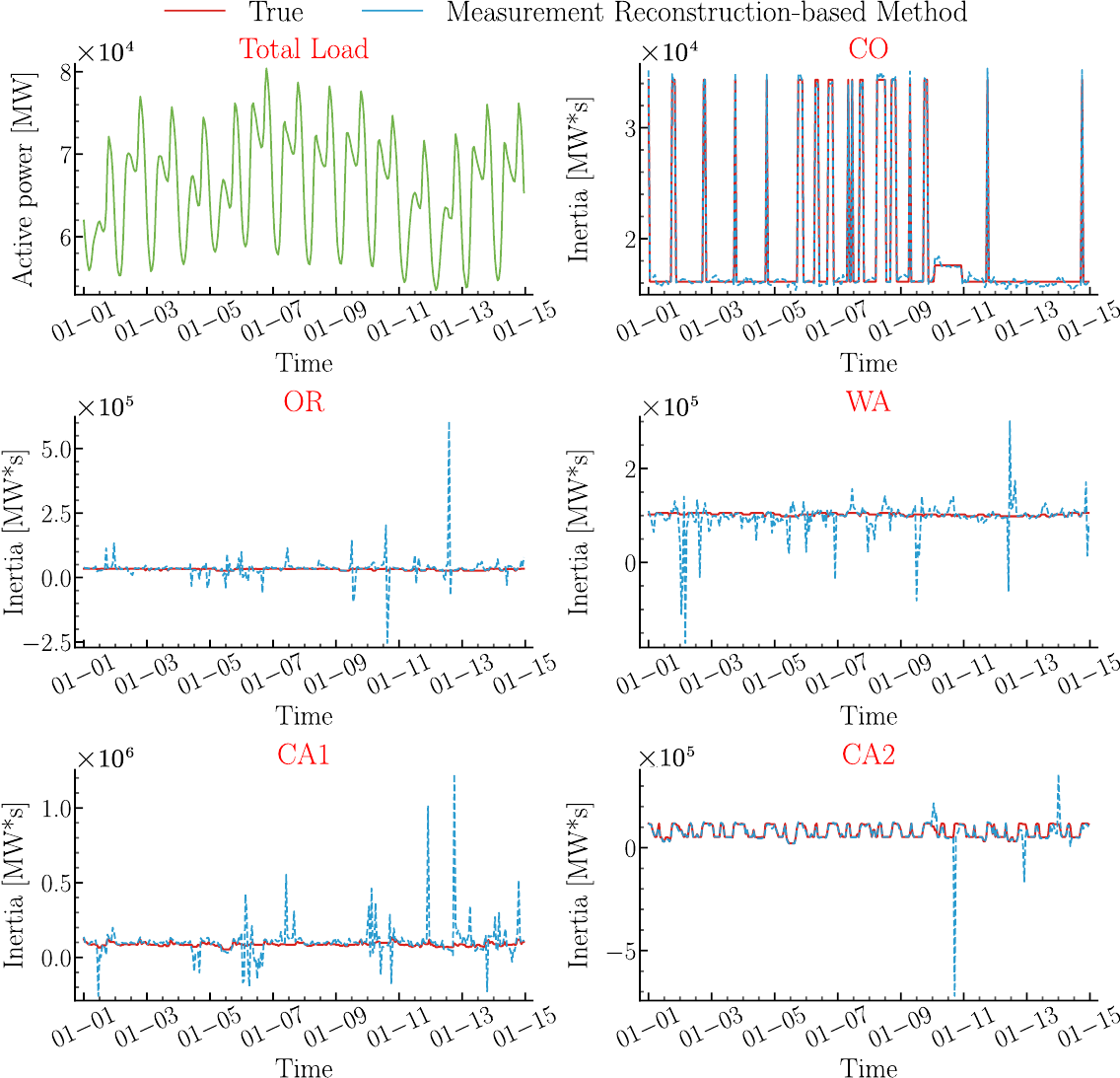}
\caption{Inertia estimation results of measurement reconstruction-based method for each area from January 1st to January 14th, 2019.}
\label{fig_DMD_inertis_est}
\end{figure}
This study extends the validation of both the system identification and measurement reconstruction-based methods across diverse operational conditions. The two-week hourly operational data, spanning January 1-14, 2019, was obtained using NREL's Multi-timescale Integrated Dynamic and Scheduling Model (MIDAS)\cite{Tan_2023}. For each operational condition, a consistent disturbance and measurement noise, as previously described, is introduced. Owing to page constraints, we present the estimation results exclusively for states CO, OR, WA, CA1, and CA2, as these exhibit notable inertia fluctuations over time, as shown in Figs. \ref{fig_transfer_inertis_est} and \ref{fig_DMD_inertis_est}. From the data, it is evident that the total system load undergoes periodic fluctuations, leading to the disconnection or connection of generators. As a result, the total inertia in each area exhibits time-varying characteristics. The system identification-based method showcases consistent estimation performance, with the maximum estimation error remaining below 15\%, as depicted in Fig. \ref{fig_transfer_inertis_est}. While its estimation outcomes tend to be lower than the actual values, indicating a biased approach, its performance is nonetheless reliable. This further illustrates its suitability for regional power systems where generator coherency might not necessarily be maintained. Conversely, due to the influence of operational conditions on the initial value \(\boldsymbol{b}\), the measurement reconstruction-based method registers errors exceeding 100\% for states OR, WA, CA1, and CA2, even though its estimations for CO appear accurate, as shown in Fig. \ref{fig_DMD_inertis_est}. This underscores the limited robustness of the measurement reconstruction-based method in estimating area inertia. In summary, the system identification-based method emerges as the more favorable approach for practical power system area inertia estimation.
\vspace{-0.3cm}
\section{Conclusions}
In this paper, we undertook an exhaustive comparative analysis to assess the efficacy of system identification, measurement reconstruction, and electromechanical oscillation-based methods for area inertia estimation in a large-scale practical power system, specifically the WECC 240-bus power system. Numerical evaluations under conditions of noise and time-varying operations underscore that the system identification-based method offers superior robustness and accuracy in area inertia estimation, despite its slight estimation bias,  compared to the other methods.

\end{document}